\documentclass{ws-procs9x6}

\def\th{\theta}

\def\si{\sigma}

\def\ta{\tau}

\def\to{\rightarrow}

\def\beq{\begin{eqnarray}}
\def\eeq{\end{eqnarray}}

\def\frac#1#2{\textstyle{{{#1} \over {#2}}}}

\def\lsim{\mathrel{\rlap{\lower4pt\hbox{\hskip1pt$\sim$}}
    \raise1pt\hbox{$<$}}}
\def\gsim{\mathrel{\rlap{\lower4pt\hbox{\hskip1pt$\sim$}}
    \raise1pt\hbox{$>$}}}

\def\aL{(a_L)}
\def\cL{(c_L)}

\def\nh^#1{{\hat N}^{#1}}

\def\nuebar{\bar\nu_e}
\def\numubar{\bar\nu_\mu}
\def\nutaubar{\bar\nu_\tau}

\begin{document}

\title{TESTING LORENTZ SYMMETRY WITH THE DOUBLE CHOOZ EXPERIMENT}

\author{TEPPEI KATORI AND JOSHUA SPITZ}

\address{Laboratory for Nuclear Science,\\
Massachusetts Institute of Technology,\\
Cambridge, MA, 02139, USA\\
E-mail: katori@fnal.gov}

\begin{abstract}
The Double Chooz reactor-based oscillation experiment 
searches for an electron antineutrino disappearance signal to investigate 
the neutrino mass matrix mixing angle $\th_{13}$. 
Double Chooz's reported evidence for this disappearance is generally interpreted as mass-driven mixing through this parameter. 
However, the electron antineutrino candidates collected by the experiment can 
also be used to search for a signature of the violation of Lorentz invariance.

We study the sidereal time dependence of the antineutrino signal rate and 
probe Lorentz violation within the Standard-Model Extension (SME) framework. 
We find that the data prefer the sidereal time independent solution, 
and a number of limits are applied to the relevant SME coefficients, 
including the first constraints on those associated with Lorentz violation in the $e-\ta$ mixing sector.

\end{abstract}

\bodymatter

\section{Double Chooz}

The neutrino Standard Model can successfully describe all precision neutrino oscillation measurements to date. 
Knowledge of the mixing angle $\th_{13}$ represents the last requirement before a measurement of CP violation in the lepton sector can proceed. 
A measurement of $\th_{13}$ is therefore critical, 
and a worldwide effort in the form of short-baseline reactor-based~\cite{reactor} and long-baseline 
accelerator-based~\cite{accelerator} experiments
has been undertaken to accomplish this goal. 

The Double Chooz reactor antineutrino experiment employs 
two 4.25~GW reactors in the north of France
near the border with Belgium as an antineutrino source. 
A liquid-scintillator-based far detector, located about 
1050~m southeast of the cores, is used to detect the antineutrinos. 

We briefly describe the main features of the Double Chooz far detector below. 
Details of the Double Chooz experiment  can be found elsewhere~\cite{DC_PRD}. 
Double Chooz is designed to detect the interaction of reactor electron antineutrinos 
with free protons via inverse beta decay ($\nuebar+p\to e^++n$) resulting in the coincidence 
of a fast positron annihilation and delayed neutron capture 
on a gadolinium or hydrogen nucleus~\cite{DC_hydrogen}.

The detector is made up of four layers of concentric cylinders. 
The innermost region is the antineutrino target volume, 
where a 10.3~m$^3$ acrylic tank is filled with gadolinium-doped liquid scintillator. The next layer is the ``gamma-catcher," 
where a 55~cm thick volume of liquid scintillator is 
used to fully reconstruct gamma rays originating in the antineutrino target region. 
The third layer is a 105~cm thick mineral oil buffer, 
where 390~10-inch PMTs are located. 
Then, after a stainless steel wall, there is a 50~cm thick inner veto region, where 78~8-inch PMTs are installed in 
the liquid scintillator to detect particles originating from the outside. 
The dominant backgrounds are spallation products, 
stopping muons, and fast neutrons. 
However, these backgrounds are directly constrained 
from the Double Chooz reactor-off data~\cite{DC_off}.
 

\section{Sidereal variation analysis}

\begin{figure}[ht]
\centerline{\epsfxsize=4.0in\epsfbox{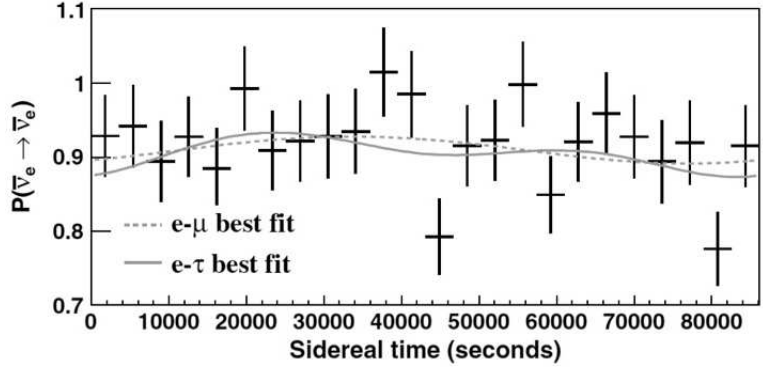}}   
\caption{The electron antineutrino disappearance probability as a function of sidereal time 
overlaid with the best fit curves~\cite{DC_LV}. 
The solid curve ($e-\ta$ best fit) assumes all Lorentz violating oscillations 
occur in $\nuebar\to\nutaubar$,   
and the dashed curve ($e-\mu$ best fit) assumes all Lorentz violating oscillations 
occur in $\nuebar\to\numubar$. Both fits are dominated by the sidereal time independent terms. 
\label{fig:LVfit}}
\end{figure}

Double Chooz has excluded the no-oscillation hypothesis at the 2.9$\si$ level~\cite{DC_PRD}. 
The analysis reported here is based on the same data set and 
is used to look for a sidereal time dependence among 
the 8249~antineutrino-induced inverse beta decay candidates. 
An observed sidereal time dependence of an experimental observable is widely considered a smoking gun of Lorentz violation. 
The analysis is performed under the SME formalism~\cite{Mewes_first}. 
The relatively small observed oscillation signal allows the effective Hamiltonian to be expanded with each oscillation channel written 
as one matrix element~\cite{Mewes_SBA}. The disappearance can be written in terms of two oscillation channels, assuming 
there are no neutrino-antineutrino oscillations:
\beq
P(\nuebar\to\nuebar)=1-P(\nuebar\to\numubar)-P(\nuebar\to\nutaubar)
\eeq
This allows access to the $e-\ta$ sector of SME coefficients for the first time. 
The disappearance signal is found to be compatible with the time-independent solution (Figure~\ref{fig:LVfit}) 
and limits on combinations of SME coefficients in the $e-\mu$ and $e-\ta$ sectors are extracted~\cite{DC_LV}. 

\section{SME coefficient limits}

Using the limits reported in the Double Chooz analysis~\cite{DC_LV}, 
we extract the upper limits on each individual SME coefficient. 
Table~\ref{tab:SME} shows the results. 

\begin{table}
\tbl{The extrapolated SME coefficients from the Double Chooz analysis. 
We use 4.2~MeV as the average antineutrino energy.}
{
\begin{tabular}{ccc}
\hline
\hline
SME coefficients & $e-\ta$ fit & $e-\mu$ fit \\
\hline
Re$\aL^T   $ or Im$\aL^T   $&7.8$\times 10^{-20}$~GeV&---                     \\
Re$\aL^X   $ or Im$\aL^X   $&4.4$\times 10^{-20}$~GeV&1.6$\times 10^{-21}$~GeV\\
Re$\aL^Y   $ or Im$\aL^Y   $&9.0$\times 10^{-20}$~GeV&6.1$\times 10^{-20}$~GeV\\
Re$\aL^Z   $ or Im$\aL^Z   $&2.7$\times 10^{-19}$~GeV&---                     \\
Re$\cL^{XY}$ or Im$\cL^{XY}$&3.4$\times 10^{-18}$    &---                     \\
Re$\cL^{XZ}$ or Im$\cL^{XZ}$&1.8$\times 10^{-17}$    &---                     \\
Re$\cL^{YZ}$ or Im$\cL^{YZ}$&3.8$\times 10^{-17}$    &---                     \\
Re$\cL^{XX}$ or Im$\cL^{XX}$&3.9$\times 10^{-17}$    &---                     \\
Re$\cL^{YY}$ or Im$\cL^{YY}$&3.9$\times 10^{-17}$    &---                     \\
Re$\cL^{ZZ}$ or Im$\cL^{ZZ}$&4.9$\times 10^{-17}$    &---                     \\
Re$\cL^{TT}$ or Im$\cL^{TT}$&1.3$\times 10^{-17}$    &---                     \\
Re$\cL^{TX}$ or Im$\cL^{TX}$&5.2$\times 10^{-18}$    &---                     \\
Re$\cL^{TY}$ or Im$\cL^{TY}$&1.1$\times 10^{-17}$    &---                     \\
Re$\cL^{TZ}$ or Im$\cL^{TZ}$&3.2$\times 10^{-17}$    &---                     \\
\hline
\hline
\end{tabular}
\label{tab:SME}}
\end{table}

The fit is done separately by assuming $\nuebar\to\nutaubar$ oscillation only or $\nuebar\to\numubar$ oscillation only. 
Therefore, there are two sets of relevant SME coefficient limits. 
With regard to the latter case, since a Lorentz violation analysis with the MINOS near detector~\cite{MINOS_LV} reports an 
order of magnitude higher-sensitivity for $\cL^{\mu\nu}_{e\mu}$ due to their higher energy beam, 
we set these coefficients to zero. 
Therefore, we do not report limits on $\cL^{\mu\nu}_{e\mu}$. 
Also, there is no time-independent limit in the $e-\mu$ sector reported 
since Double Chooz does not provide a limit on the time-independent amplitude. 

To extract each limit, we set all SME coefficients but one to zero. 
In this way, the real part and imaginary part have the same constraints. 
With this study, all oscillation channels have associated SME coefficient constraints, and it can now be considered challenging to discover Lorentz violation with a terrestrial-based neutrino experiment. 
However, there is still room to search for Lorentz violation with neutrinos. 
That is, there are manifestations of Lorentz violation which 
do not affect neutrino mixing and cannot be constrained with neutrino oscillation experiments. 
The most famous example of this is neutrino time-of-flight measurements~\cite{Mewes_highD}. 
Also, if we admit Lorentz violation is only a second-order effect of neutrino oscillations, behind neutrino mass, 
a perturbative approach in searching for a small Lorentz violation may be more efficient~\cite{Diaz_LBA}.

In conclusion, a search for Lorentz violation has been 
performed using the Double Chooz data. 
No evidence for this process has been observed.
The results are used to extract limits on the relevant SME coefficients, 
and Lorentz violation in the $e-\ta$ mixing sector are constrained for the first time. 
With the addition of this analysis amongst the world's data, 
Lorentz violation has been tested in all oscillation channels.

\end{document}